\documentclass[useAMS,usenatbib]{mn2e}

\usepackage{xspace}
\usepackage{graphicx}

\usepackage{color}
\usepackage{ulem}

\newcommand{\mycomm}[1]{#1}

\newcommand{\phiob}{$\Phi_{\mathrm{OB}}$\xspace}

\newcommand{\phiholmes}{$\Phi_{\mathrm{HOLMES}}$\xspace}
\newcommand{\phitot}{$\Phi_{\mathrm{total}}$\xspace}

\newcommand{\msun}{~M$_{\odot}$}

\newcommand{\lsun}{~L$_{\odot}$}

\newcommand{\cmcub}{~cm$^{-3}$}

\newcommand{\qh}{$Q({\rm{H^{0}}})$}

\newcommand{\Ha}{H$\alpha$\xspace}
\newcommand{\Hb}{H$\beta$\xspace}
\newcommand{\hi}{H~{\sc i}}
\newcommand{\hii}{H~{\sc ii}}

\newcommand{\Hei}{He~{\sc i} $\lambda$5876}

\newcommand{\Nii}{[N~{\sc ii}] $\lambda$6584}
\newcommand{\nii}{[N~{\sc ii}]}
\newcommand{\Oi}{[O~{\sc i}] $\lambda$6300}
\newcommand{\oi}{[O~{\sc i}]}
\newcommand{\Oii}{[O~{\sc ii}] $\lambda$3726 + $\lambda$3729}
\newcommand{\oii}{[O~{\sc ii}]}
\newcommand{\Oiii}{[O~{\sc iii}] $\lambda$5007}

\newcommand{\oiii}{[O~{\sc iii}]}

\newcommand{\Sii}{[S~{\sc ii}] $\lambda$6716}
\newcommand{\sii}{[S~{\sc ii}]}

\newcommand{\Hep}{He$^{+}$}

\newcommand{\Opp}{O$^{++}$}

\newcommand{\Nepp}{Ne$^{++}$}

\def\aap{A\&A}

\def\aj{AJ}

\def\apjl{ApJL}
\def\apjs{ApJS}
\def\apj{ApJ}

\def\araa{Annual Review of Astron and Astrophys}

\def\mnras{MNRAS} 
\def\nat{{Nature}}              

\def\nt/f{Nuclear Technology/Fusion}

\def\pasp{Public. of the Astron. Soc. Pac.}

\title[Ionization of the diffuse gas in galaxies]{Ionization of the diffuse gas in galaxies : \\
 Hot low-mass evolved stars at work}

\author[Flores-Fajardo et al.]{N. Flores-Fajardo$^1$\thanks{E-mails:NahieFlores@Gmail.com, Chris.Morisset@Gmail.com, grazyna.stasinska@obspm.fr},
  C. Morisset$^{1,2}$, G. Stasi\'nska$^2$, L. Binette$^1$\\
$^1$Instituto de Astronom\'{\i}a,
     Universidad Nacional Aut\'onoma de M\'exico\\
     Apdo. postal 70--264; Ciudad Universitaria;
     M\'exico D.F. 04510; M\'exico.\\
$^2$LUTH, Observatoire de Paris, CNRS, Universit\'e Paris Diderot; Place Jules Janssen 92190 Meudon, France.}

\begin{document}

\date{Accepted ***}

\pagerange{\pageref{firstpage}--\pageref{lastpage}} \pubyear{2011}

\maketitle

\label{firstpage}

\begin{abstract}
We revisit the question of the ionization of the diffuse medium in late type galaxies, by studying NGC 891, the prototype  of edge-on spiral galaxies. The most important challenge for the models considered so far was the observed increase of \oiii/\Hb, \oii/\Hb, and \nii/\Ha\ with increasing distance to the galactic plane. We propose a scenario based on the \textit{expected }population of  massive OB stars and hot low-mass evolved stars (HOLMES) in this galaxy to explain this observational fact. 
In the framework of this scenario we construct a finely meshed grid of photoionization models. For each value of the galactic latitude $z$ we look for the models  which simultaneously fit the observed values of the \oiii/\Hb, \oii/\Hb, and \nii/\Ha  ratios.  For each value of $z$ we find a range of solutions  which depends on the value of the oxygen abundance. The models  which fit the observations indicate a systematic decrease of the electron density with increasing $z$. They become dominated by the HOLMES  with increasing $z$ only when restricting to solar oxygen abundance models, which argues that the metallicity above the galactic plane should be close to solar. They also indicate that N/O increases with increasing $z$.
\end{abstract}



\section{Introduction}
\label{sec:intro}

In 1963, \citeauthor{1963AuJPh..16....1H} suggested the existence of an ionized layer about the Galactic plane, to explain the radio frequency spectrum observed in directions towards the Galactic pole. Assuming an electron temperature of $10^4$~K, they estimated the electron density to be $\sim$ 0.1\cmcub\ and the total mass of  ionized gas  to be $\sim 5 \times 10^8$\msun. The discovery of pulsars \citep{1968Natur.217..709H} and the use of their dispersion measure to estimate the mean electron density along the path \citep{1969A&A.....3..347R} indicated the existence of extended zones of ionized gas in the Galaxy. The detection of faint optical line emission  outside the classical Galactic \hii\ regions \citep{1971PhDT.........1R} clearly established the existence of a widespread diffuse ionized medium in the Galaxy. However, it was only after the discovery of pulsars in globular clusters located far from the Galactic plane that the scale height of this warm ionized medium (WIM) started being really appreciated:  \cite{1989ApJ...339L..29R} showed that the ionized gas extends as far as 4~kpc from the Galactic plane and estimated a free-electron scale height of about 1.5~kpc. The detection of extra-planar ionized gas in edge-on spiral galaxies using deep \Ha\ imaging \citep{1990A&A...232L..15D,1999ApJ...522..669H} enabled a better description of the vertical distribution of this  extraplanar diffuse ionized gas (eDIG) which permeates the thick disks and haloes of the galaxies.

The DIG turns out to be a major component of the interstellar medium in galaxies \citep{1991IAUS..144...67R} and investigating its topology, composition, ionization conditions and relation to the other phases of the interstellar medium -- cold neutral medium (CNM), warm neutral medium (WNM) and hot ionized medium (HIM) -- is needed for a complete understanding of the evolution of the interstellar medium in galaxies and its impact in the intergalactic medium. 

Most specialists agree that massive OB stars in  galaxies likely represent the main source of ionizing photons for the eDIG \citep[see][and references therein]{2009RvMP...81..969H}. However, the fact that the eDIG is detected well above the thin disks of galaxies implies some porosity of the interstellar medium, so that part of the ionizing photons emitted by the OB stars, whose scale height is $\sim$ 100 pc or less, arrive unabsorbed at heights of the order of a kiloparsec or more \citep{2003ApJ...586..902H}. Alternatively, an additional source of ionization could contribute at such latitudes. The existence of an additional ionizing source is suggested by the reported increase of such emission line ratios as  \nii/\Ha, \sii/\Ha, and \oiii/\Hb with galactic height, which cannot be reproduced with models of photoionization by hot, massive stars, even taking into account the hardening of the ionizing radiation due to intervening absorption (see \citealt{2003ApJ...586..902H} and references therein). The sources of additional ionization/heating that are most commonly invoked are shocks \citep{2001ApJ...551...57C},  turbulent mixing layers \citep{1993ApJ...407...83S,2009ApJ...695..552B}, magnetic reconnection, cosmic rays, or photoelectric emission from small grains \citep{1999ApJ...525L..21R}. \mycomm{Note that the need of additional heating sources has also be discussed in another context, that of ionized gas in the disks of spiral galaxies beyond the \hi disk \citep{1997ApJ...490..143B}.}

Surprisingly, the role of hot low-mass evolved stars (from now on referred to as HOLMES), which are plentiful in the thick disks and lower haloes of galaxies, and have been recognized long ago to be a significant source of UV photons in galaxies \citep{1972A&A....17..155H,1973ApJ...181..115R,1974ApJ...193...93T,1975ApJ...201..168L} were reconsidered for the eDIG only by \cite{1991PASP..103..911S} but seem to have been disregarded since then. However, it has been convincingly shown that the populations of the HOLMES expected to be found in galaxies can explain the observed line intensities in early-type galaxies \citep{1994A&A...292...13B, 2008MNRAS.391L..29S}. These stars produce a much harder radiation field than massive OB stars, and are therefore able to heat the eDIG to higher electron temperatures.

The purpose of this paper is to show that the expected populations of  HOLMES, when combined to the population of massive OB stars, are perfectly able to reproduce the distribution of emission line ratios in the eDIG.  We chose to focus on one object: the edge-on spiral galaxy NGC 891,  a galaxy located at a distance of 10 Mpc, with overall properties similar to those of the  Milky Way \citep{1984A&A...140..470V}, that has been extensively observed, especially in optical emission lines, providing the most demanding diagnostics for our scenario. 

The organization of the paper is as follows. Section~\ref{sec:obs} presents the observational data to be explained. Section~\ref{sec:scen} describes the scenario we explore. Section \ref{sec:mod} explains how we computed the flux and spectral energy distribution  of the radiation from the massive OB stars and from the HOLMES and describes our grid of photoionization models using combinations of those as an input. Section~\ref{sec:results} presents the confrontation of our models with the observational data. Finally Section~\ref{sec:conclusions} summarizes our results and presents prospects for future work. The appendix discusses several effects not taken into account in our principal grid of models, showing that they do not alter our main conclusion, i.e. that the expected population of HOLMES can explain the emission lines observed in the eDIG.

\section{Observational Data}
\label{sec:obs}

\begin{figure*}
\includegraphics[width=17cm]{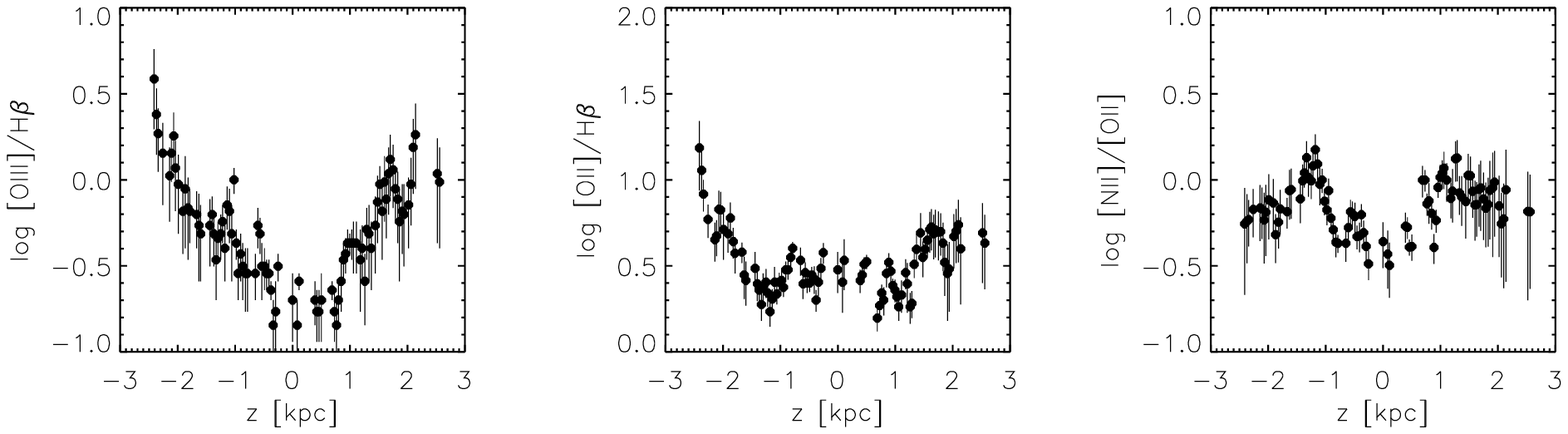}
\caption{Observed values of \oiii/\Hb, \oii/\Hb, and \nii/\oii in NGC 891, corrected for reddening,  as a function of the distance $z$ to the galactic plane. The data are from \citet{2001ApJ...560..207O}.}\label{fig:obser}
\end{figure*}

Quite a number of spectroscopic observations of the extraplanar gas in NGC 891 have been accumulated over the years \citep{1990IAUS..144P..43D,1992A&A...254L..25D,1994ApJ...423..190P,1998ApJ...501..137R,2001ApJ...560..207O,2008ApJ...680..263R} as compiled by \cite{2009RMxAA..45..261F} for the  DIGEDA\footnote{DIGEDA (Diffuse Ionized Gas Emission Data base) is available in Vizier website: http://vizier.u-strasbg.fr/cgi-bin/VizieR.} data base.
The most interesting data set from our point of view is the one by \cite{2001ApJ...560..207O}, in which the intensities of \Ha, \Hb, \nii, \oii, \oiii, and \sii\footnote{Unless explicitly stated otherwise, in the whole paper \nii, \oi, \oii, \oiii, and \sii\ stand for \Nii, \Oi, \Oii, \Oiii, and \Sii, respectively.} were measured along a slit placed perpendicularly to the plane of the galaxy. The availability of the  \Hb\ line allowed the authors to correct the observed line ratios for reddening, assuming an intrinsic value 
of \Ha/\Hb\, of 2.9. Care was taken by the authors to account for stellar absorption in the \Hb\ line  as well as possible. The most important advantage of the data set by \cite{2001ApJ...560..207O} with respect to previous data sets is that it contains the \oii\ line, which provides an important diagnostic of the ionization conditions in the eDIG, when complemented with the \oiii\ and hydrogen lines. In absence of \oii\ one is forced to rely on either \nii\ or \sii\ to estimate the excitation level of the gas. However the use of \nii\ implies making an assumption on N/O, while nothing is a priori known  on this abundance ratio within the eDIG. Regarding \sii, while the S/O ratio can be expected to be identical to the one in the Sun or in classical \hii\ regions, the use of this low excitation line in conjunction with \oiii\ may lead to spurious interpretations:  this line is emitted in a zone of very low excitation and it is known that, even in classical \hii\ regions, photoionization models in which the gas distribution is homogeneous fail to simultaneously reproduce the intensities of \oiii, \oii, \nii, \sii\ with respect to the hydrogen lines \citep[see e.g.][]{2006MNRAS.371..972S, 2010AJ....139..712L}.

Figure \ref{fig:obser} reports the intensity ratios \oiii/\Hb, \oii/\Hb, and \nii/\oii\, as a function of $z$, the distance to the plane of the galaxy. In absence of tables giving the numbers in the original paper, the figures of \cite{2001ApJ...560..207O} were digitized and the line ratios converted to the ones we show in Figure \ref{fig:obser}, which will be directly compared to our models. We also digitized and converted the error bars, which are shown in our plots as vertical lines\footnote{ The error bars were obtained from the errors plotted in Fig. 6a of \cite{2001AAS...19911705O} by assuming that the errors in \oiii/\Hb and \oii/\Hb were 2.9 times those in \oiii/\Ha and \oii/\Ha respectively, and that those in \nii/\oii were the quadratic sum of the error in \nii/\Ha and oii/\Ha.}$^,$\footnote{The digitized  data from \cite{2001ApJ...560..207O} and the data for the error bars were added to an updated version (V2.0) of DIGEDA. In this new version we corrected a confusion between \sii $\lambda$6717 + 6731 and \sii $\lambda$6717 and  added a code,  OBS\_ID, for the identification of each error with each observation ($\rm OBS\_ID_{error}$=$\rm OBS\_ID_{data}$ + 100000).}. Note that these error bars  reflect only the errors of the flux measurements. In the original figure of \citeauthor{2001AAS...19911705O}, there are several spikes in \oii\ in the region of the plane of the galaxy. The authors warn that it is not clear how physical those peaks are. We have therefore removed the corresponding points in our figure. 
As has been noted by several authors, the \oiii/\Hb ratio increases steadily with $|z|$. The \oii/\Hb ratio also shows this tendency, although less strongly. The  \nii/\oii\ shows a shape in ``M'', strongly suggesting that the N/O ratio is not constant. Ignoring this fact when interpreting the data in terms of the ionization state of the DIG would then lead to spurious results. 
Note that part of the light from \hii regions in the disk could be dust-scattered outside the disk and contribute to the extra planar emission. Since the albedo is almost independent of  wavelength in the optical the
line ratios of the scattered component should be  similar to the ones near the galactic plane.
The fact that the extra planar emission has different line ratios implies that the scattered component is not dominant.

\citet{1998ApJ...501..137R} also observed the faint \oi\, line (with a slit not exactly at the same position as that of \citealt{2001ApJ...560..207O}) and plotted it as a function of $z$. Since the \oi\ line has a similar problem as \sii\, when attempting to interpret observations with the help of homogeneous photoionization models, we will not use this observation to constrain our models. However, once our fitting procedure is achieved, we will compare our predictions with those observations (see Sect \ref{sec:results}). The same will be done with the  \Hei\, line, extracted from Fig. 3 of \cite{2008ApJ...680..263R}.

\citet{2008ApJ...680..263R} and \citet{2011arXiv1101.1491R} reported mid infra-red spectroscopy of the diffuse ionized halo in NGC 891, which showed that the [Ne~{\sc iii}] $\lambda$15.6$\mu$m/[Ne~{\sc ii}] $\lambda$12.8$\mu$m is enhanced in the extraplanar pointings relative to the disk pointing. We will also confront our models to these observations (see Sect \ref{sec:results}). 

Finally, observations of the vertical distribution of the \Ha\ surface brightness given by \cite{1998ApJ...501..137R} and   \cite{2003ApJS..148..383M}  provide important information on the number of ionizing photons absorbed as a function of height above the galactic plane.

\section{Scenario and modelling policy}
\label{sec:scen}

\begin{figure}
\includegraphics[angle=0,width=8cm]{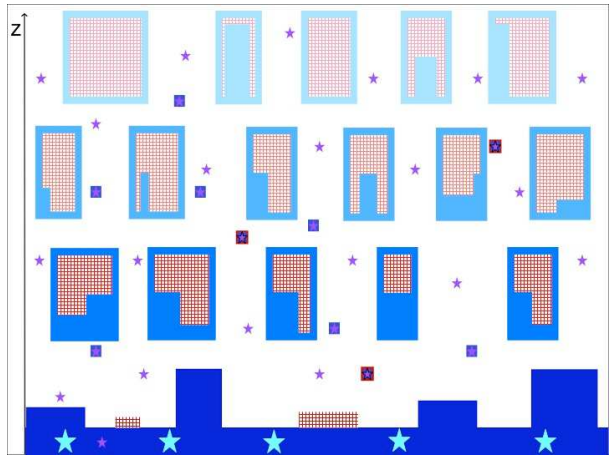}
\caption{Schematic representation of the extraplanar gas and ionizing stars. Ionized gas is represented in blue, neutral gas in hashed-red, with a lighter shade for less dense regions. See text for a detailed description.}\label{fig:scenario}
\end{figure}

We first present a qualitative view of the scenario that we propose to explain the emission line features of the DIG.
This is shown in Fig. \ref{fig:scenario}, where the galactic plane is located at the bottom of the figure. 
The extraplanar gas is distributed in clouds, between which ionizing photons emitted by the massive stars located in the disk of the galaxy can travel up to high latitudes. For simplicity, we represent the gas clouds by rectangles. 
Ionized gas is represented in blue while neutral gas is represented in red. For both components, a darker shade indicates  higher density gas. All that is white represents very diffuse, coronal gas which is neither emitting in the optical nor absorbing. The big blue stars in the plot  represent massive OB stars. They are located close to the galactic plane, and surrounded by \hii\ regions, some of which are density-bounded and allow leakage of ionizing photons. The small purple stars represent the HOLMES, which can be either central stars of present-day planetary nebulae, or hot pre-white dwarfs whose evolution time-scale is larger than that of the nebular envelope they ejected in a previous stage. These stars are much hotter than massive OB stars. They are also much more numerous and their height-scale is that of the old stellar population in the galaxy.  Most planetary nebulae, although surrounded by a gaseous envelope of density $n_{\rm e} \sim 10^{2}-10^{4}$\cmcub\ are density-bounded and allow leakage of hard ionizing photons in the diffuse medium; some of them, however, are ionization-bounded \mycomm{\citep[see e.g. Fig 3 from][which shows the expected proportion of optically thick and thin planetary nebulae among a simulated sample of bright planetary nebulae]{2004A&A...423..995M}}. In both cases, the ionizing photons produced by the central stars are ultimately absorbed by gas -- be it the proper planetary nebula or the diffuse interstellar gas -- and, at the spatial resolution of the observations, they can be counted as photons available to ionize the eDIG. More importantly, most of the HOLMES are not surrounded by a planetary  nebula, and the total number of ionizing photons they produce  is thus far greater than what can be estimated from planetary nebulae alone. 

The part of the radiation  from the massive stars embedded in the classical \hii regions escapes  and, in a first approximation, ionizes the ``bottom'' parts of the gas clouds. As $z$  increases, the surface fraction of the gas clouds that are not shielded by intervening neutral zones becomes smaller. At the same time, the radiation originating from the massive stars becomes harder, since it is the photons with energies closest to the ionization threshold of hydrogen which are absorbed first. 

The HOLMES, which are distributed in the galaxy thick disk and halo, also contribute to the ionization of the clouds. Their influence with respect to that of OB stars increases away from the galactic plane, and the ``bottom'' of the clouds is ionized by a radiation field that becomes harder and harder. At any latitude, HOLMES will also produce an ionized skin on all the sides of the gaseous clouds that are not facing the OB stars.  \mycomm{The clouds have to be optically thick, i.e. to possess a neutral core, since the \oi\ line is detected at all heights above the galactic plane.} 

\begin{figure}
  \includegraphics[angle=0,width=8cm]{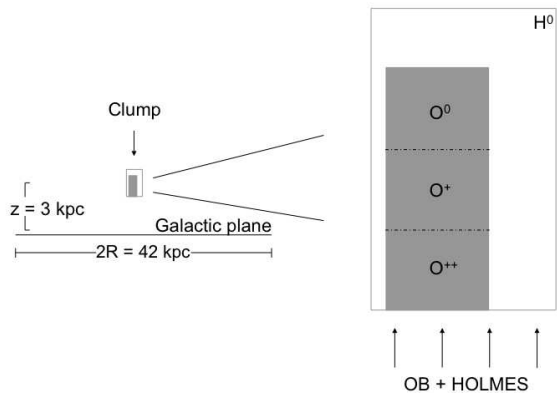}
\caption{How the photoionzation models are computed. Each cloud is photoionized by radiation coming from the galactic disk (OB stars) and from the thick disk/halo (HOLMES). The computations are done in plane parallel geometry with the ionizing source on one side only.}\label{fig:geometry}
\end{figure}

In the photoionization models presented below we consider, for simplicity, that each cloud is ionized only from the ``bottom'', by the summed radiation field from the OB stars and from the HOLMES\footnote{A short discussion of a  model with more elaborate geometry is presented in Appendix~\ref{sec:effect-geom-distr}, where we show that the main conclusions are indeed not affected.}. The models are computed in the plane parallel approximation, assuming that each cloud is of constant density. The computations are stopped when the ionization fraction of hydrogen drops below 0.1\%. Fig. \ref{fig:geometry} shows the details of the ionization structure of a given cloud. 

 \mycomm{Note that the covering factor of the whole population of clouds is not necessarily unity, and some ionizing radiation may escape from the galaxy.} 

\section{Quantitative modelling}
\label{sec:mod}

In this section, we first explain how we compute the ionizing radiation field  that reaches the clouds, then we describe the grid of photoionization models constructed to interpret the observations. 

\subsection{OB stars}
\label{sec:OB}

The ionizing flux issuing from OB stars  is obtained from the evolutionary stellar population synthesis code Starburst99 \citep{1999ApJS..123....3L}. We adopted the stellar initial mass function of \citet{1993Kroupa_mnra262} for stellar masses  ranging from 0.1 to 100\msun. The stellar tracks are the Geneva ones \citep{1994Maeder_aap287} with enhanced mass loss. The stellar model atmospheres are those of \citet{2001Pauldrach_aap375} and \citet{HM98}, implemented by \citet{2002Smith_mnra337} in Starburst99. The metallicity of the stellar population is taken to be solar. 
We considered a continuous star formation and took the radiation field at an age when it is stabilized (in practice $10^7$\,yr). We do not consider dilution of the OB radiation field, since  the latitudes we consider (up to 4\,kpc) are much smaller that the radius 
 of the stellar disk in NGC 891,  which is of $\sim$21\,kpc \citep{1984A&A...140..470V, 2005Ap.....48..221T}.

We estimate the total number of hydrogen ionizing photons, \qh$_{\rm {OB}}$, in the following way. The total infrared luminosity of NGC 891 is $L_{\rm {IR}} = 2.6 \times 10^{10}$\lsun \citep{2004ApJ...606..271G}. Using their Eq. 9 we obtain a total  present-day star formation rate of  5.2~M$_{\odot}$~yr$^{-1}$ in NGC 891
 (this assumes that the observed FIR emission originates primarily from dust heated by O, B, and A stars)\footnote{As mentioned by \citet{1998ARA&A..36..189K}, the conversion from total infrared luminosity to star formation rate is only approximate for spiral galaxies,  due to complex star formation history and geometrical issues, but only a rough estimate is sufficient for our purposes. }.  The total number of hydrogen ionizing photons for a constant stellar mass formation rate of 1~M$_{\odot}$~yr$^{-1}$  given by Starburst99 with the parameters listed above is $10^{53.18}$ ph s$^{-1}$. Thus the total number of hydrogen ionizing photons in NGC 891 is  \qh$_{\rm {OB}}$ $= 7.8 \times 10^{53}$ ph~s$^{-1}$. The vast majority of these ionizing photons is expected to be produced by OB stars located in the disk. Assuming that all of them reach the eDIG, their surface flux at the bottom of a cloud,  \phiob, would  then be given by: 
\phiob = \qh$_{\rm {OB}}$\, /(2 $\pi R^2$) $ =  3.1\times 10^7   (21/R)^{2}  $ph~cm$^{-2}$~s$^{-1}$  if the OB stars are distributed uniformly in a disk whose radius in kiloparsec is $R$. The factor 2 in the denominator accounts for the fact that the radiation produced by the OB stars  must feed both the ``upward'' and the ``downward'' directions. 

The \hii\ regions located in the disk are largely ionization bounded, and photon leakage occurs through small ``holes'', as suggested by the thin \Ha\ filaments seen perpendicular to the disk of NGC 891 \citep{2004AJ....128..674R}. Intervening optically thick eDIG clouds  further reduce the number of photons from OB stars reaching the cloud under consideration.  \mycomm{The range of values of \phiob\ explored in the models below takes into account the, a priori ill-defined, covering factor.}  An additional factor to take into account is the \textit{partial} absorption (and hardening) of the OB photons travelling in the eDIG before reaching the cloud. This is  treated in Appendix \ref{sec:effect-tau}.

\subsection{HOLMES}
\label{sec:WD}

While Starburst99 is optimized to provide a good description of the ionizing radiation field from massive stars, in particular through its  use of modern model atmospheres for OB stars, it does not compute the radiation field from post-asymptotic giant branch stars. We therefore resort to use the evolutionary spectral synthesis code PEGASE \citep{1997A&A...326..950F} with which we are able to compute the ionizing spectral energy distribution of HOLMES. Taking again a Kroupa stellar initial mass function and a solar metallicity, we consider the spectral energy distribution of a  coeval population of  stars at an age of 10 Gyr, i.e. approximately the age of old stellar populations.
The real star formation history of the old stellar populations in the halo and thick disk of NGC 891 is probably more complicated than that of the assumed instantaneous starburst. However the \textit{integrated} radiation field from all the HOLMES in an old stellar population does not depend strongly on the details of the star formation history,  as soon as ages larger than $10^8$yr are concerned. Roughly, the stellar energy distribution is equivalent to that of a $10^5$~K blackbody (thus much harder than the radiation field from massive stars), while the luminosity remains almost constant, as can be seen in Fig. 2 of \citet{2011MNRAS.tmp..249C}. For the scenario we adopted, the total number of ionizing photons of the HOLMES (obtained by integrating the spectral energy distribution from PEGASE) is $7 \times 10^{40}$ph~s$^{-1}$ \msun$^{-1}$. If we consider that the mass of the old thick disk in NGC 891 is $3 \times 10^{10}$\msun\, as inferred from \citet{1990mwg..conf...66V}, the total ionizing photon number of the HOLMES present in the old thick disk is \qh$_{\rm {HOLMES}}$ = $ 2.1 \times 10^{51}$ph~s$^{-1}$. \footnote{\mycomm{This corresponds to about $10^4$ ``typical'' planetary nebulae nuclei emitting each $2 \times 10^{47}$ph~s$^{-1}$. The  real number of HOLMES is probably significantly larger, since many of them are expected to be less luminous than a ``typical'' planetary nebulae nucleus.}}  Considering the stars located in only half of the halo, the predicted mean surface flux of the ionizing photons from the HOLMES  is \phiholmes = $ 8.4 \times 10^{4} (21/R)^2$ ph~s$^{-1}$~cm$^{-2}$, since the scale length of the thick disk is very close too that of the thin disk \citep{2009MNRAS.395..126I}.   For simplicity, we disregard the HOLMES that are present in the extended halo, so that, from this point of view,  the value of  \phiholmes that we consider is a lower limit. As for OB stars, we do not consider dilution of the radiation field. In addition, we assume that the radiation field from HOLMES is the same whatever the distance from the galactic plane, in particular we do not consider any metallicity dependence. As a matter of fact, a metallicity of one tenth of solar, rather than solar, could perhaps be more representative of the metallicity of the thick disk and halo stellar populations in NGC 891 \citep{2009MNRAS.396.1231R}, but the ionizing radiation field from the HOLMES is virtually independent of metallicity, as shown in Fig. 2 of \citet{2011MNRAS.tmp..249C}.  We also assume that the HOLMES radiation field is not modified by intervening absorption. This is a simplifying hypothesis which is justified by the fact that HOLMES are present everywhere in the eDIG. We are aware that there is much uncertainty involved in the predictions of stellar population synthesis models after the asymptotic giant phase \citep[e.g.][]{2010IAUS..262...36M}. However, we believe that our procedure should roughly give the correct energy distribution and ionizing luminosity from the integrated population of HOLMES.

\subsection{The grid of photoionization models}
\label{sec:Cloudy}

The photoionization models for the eDIG were computed with Cloudy version 07.02.01, last described by \citet{1998PASP..110..761F}. The models are computed with the plane parallel approximation.  
Each one is defined by the value of \phiob, using the unabsorbed spectral energy distribution from Starburst99, the ionization parameter $U$ and the chemical composition of the gas. So far, we have no direct information on the values of the chemical composition of the gas in the  eDIG, and there is a priori no reason to assume that it is solar\footnote{Varying the metallicity of the HOLMES or OB stellar models would be an additional refinement, however not justified by the degree of approximation of our study.}.  The abundances of the heavy elements (except N, see below) relative to O are fixed to their solar values as implemented in Cloudy (see Cloudy's manual for details). The abundances of  Mg, Si, Fe are depleted  by 1 dex. The N/O ratio, a priori, has no reason to be constant, since the nucleosynthesis sites of N and O are very different. Therefore, we also vary N/O in our models. This parameter has little impact on the \oiii/\Hb\, and \oii/\Hb\, ratios, but, of course, strongly affects the computed \nii/\oii\ ratio. 

The ionization parameter $U$ is defined by \phitot /$(n_\mathrm{e} c$), where \phitot = \phiob + \phiholmes, $n_\mathrm{e}$ is the electron density in the cloud,  and $c$ is the speed of light. The value of \phiholmes is the one given in Sect.~\ref{sec:WD}. The values of $U$ defining the models are varied by changing the density  once \phiob is fixed.  

We compute a medium resolution (MR) grid with the combinations of parameters given in Table \ref{tasb:params}. In total, this makes 3402 models, all of which are included in the 3MdB database \citep{2009MmSAI..80..397M} under the reference DIG\_MR. To improve the precision of the model-fitting for each value of $z$, we produce a high resolution (HR) grid by interpolating between the values obtain from the MR grid, dividing the size of all the steps by 2. This leads to a grid with 16 times more values than the MR grid, i.e. a total of 54432 models, referenced in 3MdB as DIG\_HR.
To speed-up computations, all the models are dust-free.  The (negligible) effect of the interaction of dust with photons and gas particles is discussed in Appendix \ref{sec:effect-dust}.
\begin{table}
\begin{center}
\caption{Parameter values for our medium resolution photoionization model grid.\label{tasb:params}}
\begin{tabular}{lrrrr}
\hline
Parameter & Min & Max & Step  & N$^a$\\
\hline
log \phiob   & 3.5  & 7.5 & 0.5 & 9\\
log O/H & -4.3 & -2.7 & 0.2 & 9\\
log N/O & -1.4 & -0.2 & 0.2 & 7\\
log $U$   & -4.0  & -3.0 & 0.2 & 6\\
\hline

\end{tabular}
\end{center}
$^a$Number of different values for each parameter.
\end{table}

\begin{figure}
\includegraphics[angle=0,width=7cm]{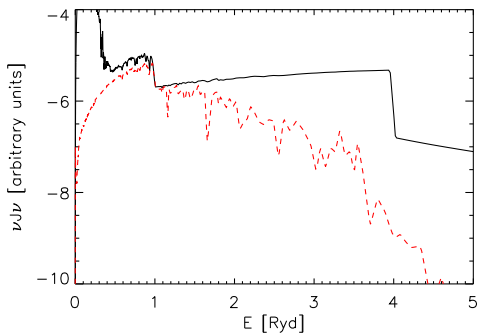}
\caption{The spectral energy distribution of the radiation emitted by HOLMES (black line) and massive OB stars (red dashed line).} \label{fig:SED}
\end{figure}

Figure \ref{fig:SED} shows the spectral energy distribution of the radiation emitted by the massive OB stars (red dashed curve) and by the HOLMES (black curves) determined as explained in Sects. \ref{sec:OB} and \ref{sec:WD}. From this, one can infer that models with larger values of \phiholmes/\phitot will have a harder ionizing radiation field and will thus produce larger \oiii/\Hb\ and \oii/\Hb\ ratios.  \mycomm{Note that the photons contributing mostly to the heating by ionization of hydrogen are between 1 and 3\,Ryd, but, in the case of HOLMES photons with energies above 4\,Ryd also contribute  to the heating, due to the photoionization of \Hep\ ions, which, in the most ionized zones,  are found in larger amounts than \textit{neutral} hydrogen particles, in spite of the He/H ratio being one tenth.} 

\section{The DIG explained!}
\label{sec:results}

\begin{figure*} 
\includegraphics[angle=0,width=17cm]{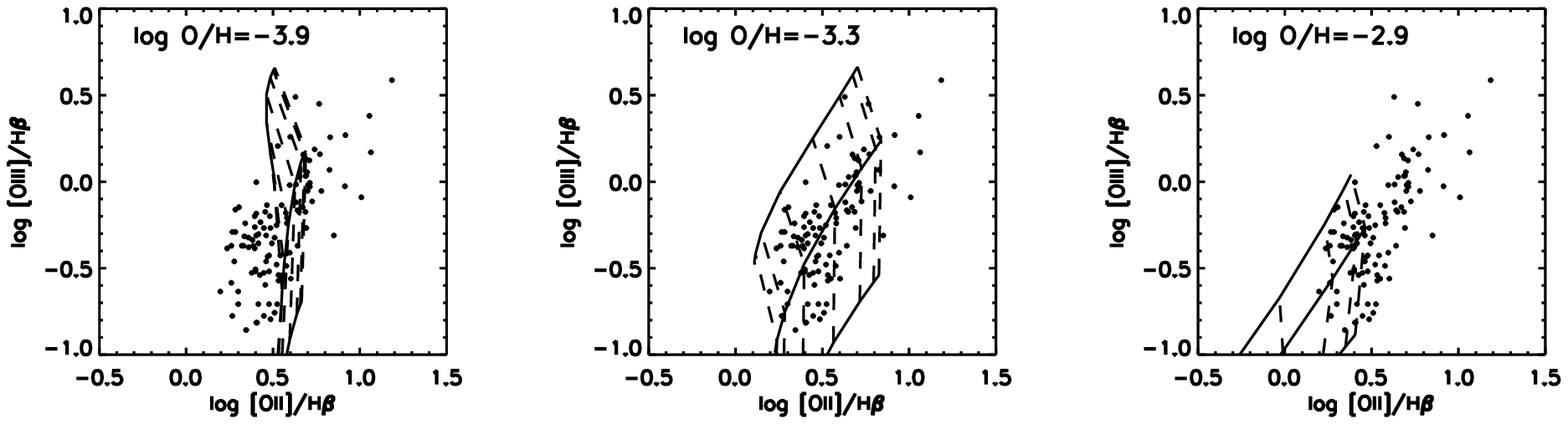}
\caption{The positions of the observational points with respect to our grid of models in the \oiii/\Hb\, vs. \oii/\Hb\, plane, for various values of O/H: from left to right, the value for log O/H is -3.9,  -3.3, and -2.9. The dashed lines join models with same values of log \phiob  (equal to 3.5, 4., 4.5, 5., 5.5, 6. 6.5), while the continuous lines join models with same values of log $U$ (equal to -4, -3.5, -3). }\label{fig:O2O3}

\includegraphics[angle=0,width=17cm]{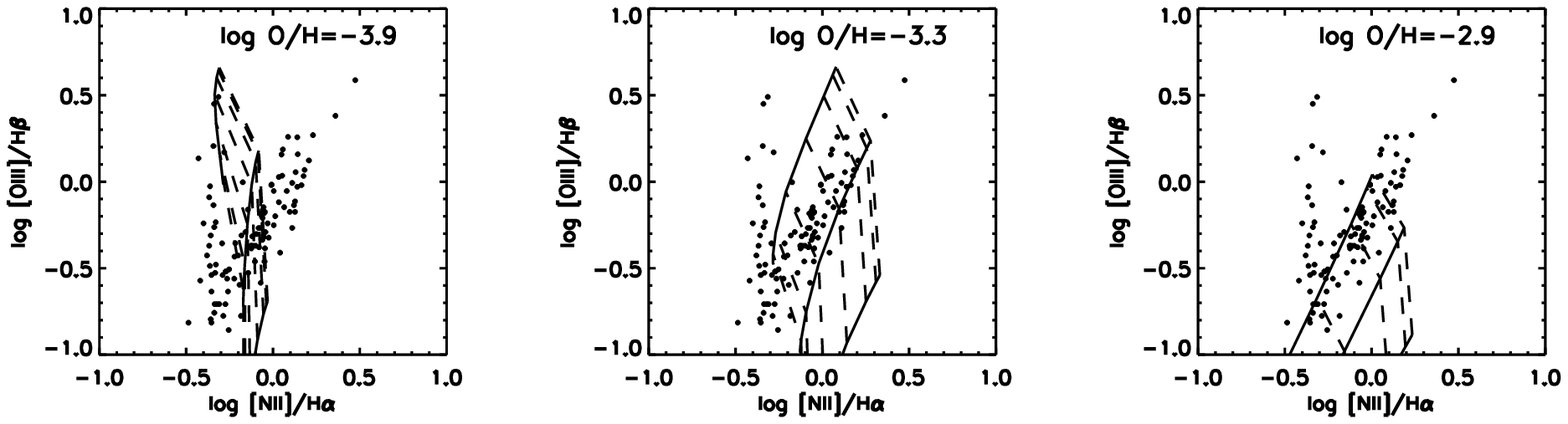}
\caption{Same as Fig. \ref{fig:O2O3} but in the \oiii/\Hb\, vs. \nii/\Ha\, plane.}\label{fig:N2O3}
\end{figure*}

Figure \ref{fig:O2O3} shows the location of the observational points in the \oiii/\Hb, vs \oii/\Hb\, diagram, with respect to our grid of models. The dashed lines correspond to models with same values of log \phiob  (ranging from 3.5 to 6.5 by steps of 0.5), while the continuous lines correspond to models with same values of log $U$ (equal to -4, -3.5, -3). Each panel corresponds to a different value of log O/H (equal to -3.9, -3.3, -2.9). The value of N/O is the same in all the panels, namely -0.5~dex (and does not strongly affect the position of the models in this plot). The observational points that have the highest values (\oiii/\Hb, \oii/\Hb) correspond to the smallest values of \phiob.  Since we do not know  the value of the metallicity in the eDIG a priori, different models can reproduce a given  couple of observed values (\oiii/\Hb, \oii/\Hb). It can be seen from these plots that the fitting models will not differ very much in terms of $U$  but can have significantly different values of \phiob, depending on O/H. There is a priori no way to discriminate between solutions with low and high metallicity\footnote{Unlike in the case of classical \hii\ regions where the degeneracy between high and low metallicities derived by strong line methods can be solved using the \nii/\oii\ or \oiii/\nii\, ratios \citep{1994ApJ...426..135M}, the nitrogen lines are of no use here, since a priori we know nothing about the N/O ratio.}. 

Figure~\ref{fig:N2O3} is the same as Fig.~\ref{fig:O2O3}, but in the (\oiii/\Hb\, vs \nii/\Ha) plane \citep[the famous BPT diagram from][]{1981PASP...93....5B}. In all the panels, we show models with N/O = -0.5~dex. It is clear that the models fitting the observed data in this figure will strongly depend on the value of N/O. This illustrates how misleading the interpretation of observed spectra can be if ignoring the possible variations of N/O.

\begin{figure*} 
\includegraphics[angle=0, ,width=17cm]{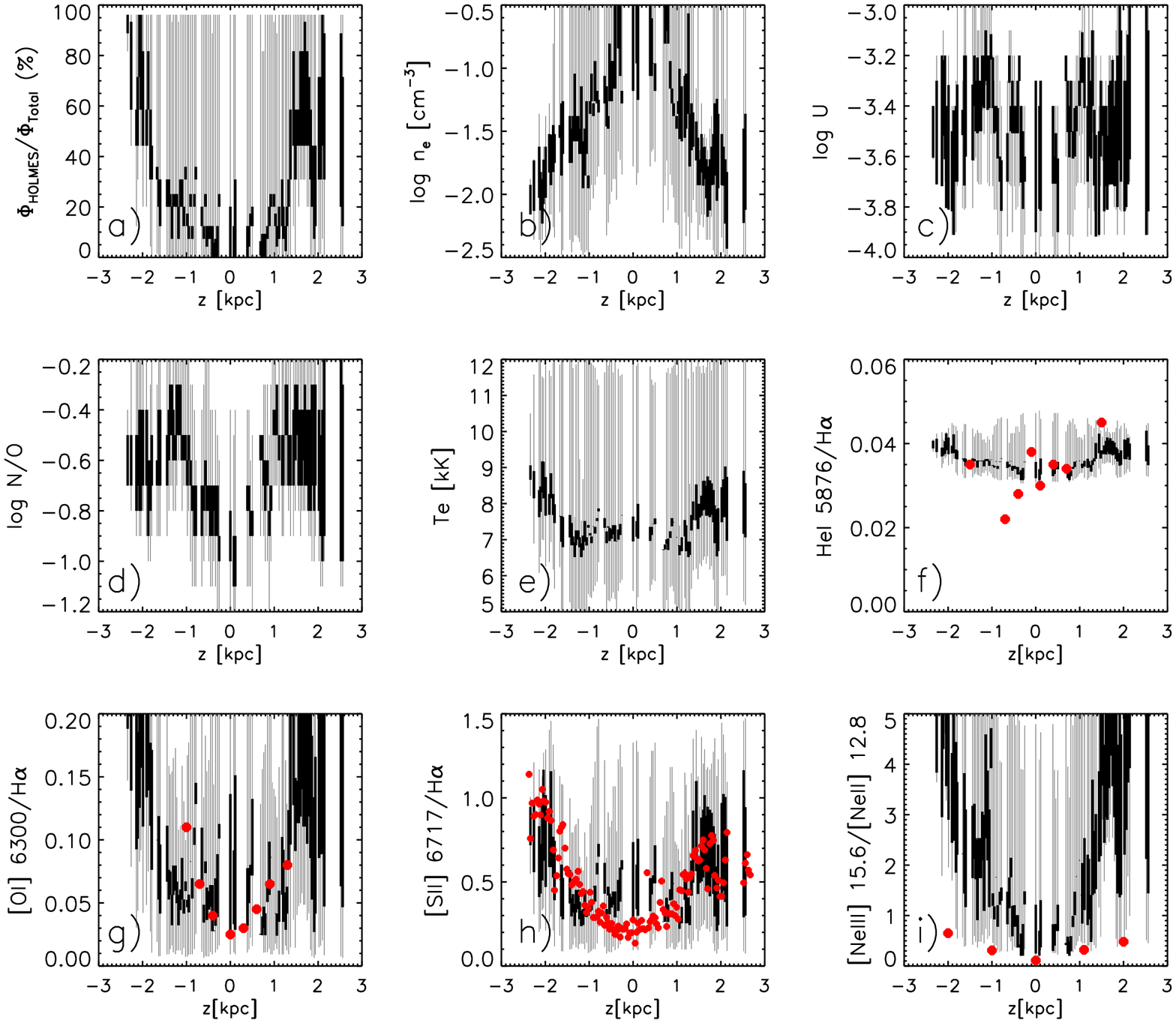}
\caption{Models that fit \oiii/\Hb, \oii/\Hb, and \nii/\Ha\ simultaneously. The light grey bars indicate the range of values for wich the models in the grid fit the observations at a given $z$. The dark bars show the same, but restricting to solar abundances models. The observed values are indicated by the big red dots. The various panels show the behaviour of various input or output parameters of the models: a) \phiholmes/\phitot; b) $n_\mathrm{e}$  (for a value of $R$ equal to 21\,kpc, see text); c) $U$; d) N/O ; e) the mean electron temperature; f) \Hei/\Ha; g)  \oi/\Ha; h) \sii/\Ha; i) [Ne~{\sc iii}] $\lambda$15.6$\mu$m/[Ne~{\sc ii}] $\lambda$12.8$\mu$m. }
\label{fig:resU}
\end{figure*}

For each observational point in the eDIG of NGC 891 we select the models of our finely meshed (HR) grid that 
\textit{simultaneously} reproduce  the values of \oiii/\Hb, \oii/\Hb, and \nii/\Ha, within the observational uncertainties depicted in Fig. \ref{fig:obser}.  

In Fig.~\ref{fig:resU} we show the acceptable ranges of values for \phiholmes/\phitot, $U$, $n_\mathrm{e}$, N/O,  \mycomm{and the mean electron temperature} as a function of $z$.  Here, $n_\mathrm{e}$ is obtained assuming $R=21$\,kpc. If a smaller value of $R$ were judged to be more appropriate, the values of the electron densities would be multiplied by $(21/R)^2$, but the other plotted quantities would remain the same. For example, adopting an effective radius of about 4.8\,kpc as being more representative of the distribution of stars in the thick disk \citep{2009MNRAS.395..126I} the values of $n_\mathrm{e}$ would range from $\sim$ 4\,cm$^{-3}$ close to thin disk to $\sim$ 0.2\,cm$^{-3}$ at $z = 2$\,kpc.  \mycomm{This change of density is a consequence of having to keep $U$ to the same value to obtain the same line ratios.}
We do not plot the acceptable range of O/H, since, for most positions, a solution can be found for a wide range of values of O/H. A broad range of possible values of \phiholmes/\phitot\ is estimated for any value of $z$ but, if we restrict to models with solar metallicity (thick black bars in Fig.~\ref{fig:resU}),  we find that this parameter tends to increases with $|z|$, implying that the role of HOLMES in the ionization and heating of the eDIG clouds becomes more important away from the plane of the galaxy.  \footnote{It is tempting to invert the argument and say that, since HOLMES are expected to play a role which is increasing with the distance to the galactic plane, the oxygen abundance in the eDIG must be rather uniform and not far from solar.} \mycomm{This produces a rise in the electron temperature as seen in Fig.~\ref{fig:resU}e.} The value of $U$  shows no clear trend, while $n_\mathrm{e}$ tends to decrease with $|z|$. This latter result is in concordance with the measured behaviour of  $n_\mathrm{e}$  as function of galactic height in the Milky Way \citep{2008A&A...490..179B}. Notice that, for simplicity, we have kept the value of \phiholmes constant, while the space density of HOLMES is expected to decrease with $|z|$, so the decrease of the electron density is likely to be even stronger. The N/O ratio as a function of $z$ shows an ``M'' shape, reminiscent of the observed \nii/\oii\, ratio, but with larger error bars. Although a constant value of N/O is marginally acceptable for the entire $z$ range sampled, we believe that the tendency we see for N/O could be real. As a matter of fact, we would expect N/O to be highest in the region where HOLMES are important, if the diluted envelopes of post-asymptotic giant branch stars, which are enriched in nitrogen through dredge-up processes in the parent stars, make a significant contribution to the eDIG.

Panels f--i of Fig. \ref{fig:resU} compares the models that fit \oiii/\Hb, \oii/\Hb, and \nii/\Ha\ simultaneously with the observations of a few other line ratios: \Hei/\Hb (observational data form \cite{2008ApJ...680..263R}), \oi/\Ha\ (data from \ \cite{1998ApJ...501..137R}), \sii/\Ha\ (data from \cite{2001ApJ...560..207O}),  and  [Ne~{\sc iii}] $\lambda$15.6$\mu$m/[Ne~{\sc ii}] $\lambda$12.8$\mu$m (data from \cite{2008ApJ...680..263R,2011arXiv1101.1491R}). The \Hei/\Hb ratio is reasonably well reproduced, given the uncertainty in the intensity of the weak \Hei line. The \sii/\Ha and \oi/\Ha ratios in the models does not reproduce the observed trends very well  especially in the regions of small $|z|$. However, as mentioned above, we cannot expect our constant-density models to reproduce the entire range of excitations.  In addition, the \oi line can be produced in photodissociation regions in the presence of a large proportion of very small grains which, as shown by \cite{2001A&A...379.1024S}, can boost the gas temperature due to photoelectric heating and therefore allow the excitation of the \oi\ line even in a neutral medium. Such a scenario was not included in the present photoionization models. Regarding [Ne~{\sc iii}] $\lambda$15.6$\mu$m/[Ne~{\sc ii}] $\lambda$12.8$\mu$m, our models overpredict this ratio.  \mycomm{It must be recalled that to produce \Nepp\ requires photons with energies of at least 3\,Ryd, i.e. much higher than needed to produce \Opp. Photoionization model predictions of the ionization structure of neon strongly rely on the number of stellar photons just above 3\,Ryd. For massive OB stars, state of the art stellar atmosphere models differ significantly in this energy range} \citep[see][]{2004A&A...415..577M,2008MNRAS.389.1009S}. For HOLMES, in addition, the predicted spectral energy distributions in this energy range strongly depend on the adopted abundances for the heavy elements, particularly of carbon. Therefore, we do not consider that the discrepancy between our models and observations as regards [Ne~{\sc iii}] $\lambda$15.6$\mu$m/[Ne~{\sc ii}] $\lambda$12.8$\mu$m disproves our models for interpreting the ionization of the eDIG.

\section{Summary and prospects}
\label{sec:conclusions}
There has been no agreement so far on the nature of the ionization sources that could explain the behaviour of the emission lines observed in the diffuse ionized gas (DIG) in late-type galaxies. Several scenarios have been proposed in the past, none of them being fully satisfactory. In this paper we have analyzed the  galaxy NGC 891, the prototype and most studied of edge-on spiral galaxies, and we proposed a model that reproduces all the considered constraints satisfactorily. The most important challenge for the models considered so far was the observed increase of \oiii/\Hb, \oii/\Hb, and \nii/\Ha\ with increasing distance from the galactic plane. 

Our model considers both the population of massive OB stars and that of hot low-mass evolved stars (HOLMES) expected to be present in the galaxy. 

The ionizing photons from the massive stars located in the galactic plane  partly escape into the thick disk and halo, where they are gradually absorbed by the clouds that constitute the diffuse medium.  The surface fraction of the gas clouds that are not shielded by intervening clouds becomes smaller with increasing distance to the galactic plane.  At large distances from the galactic plane, HOLMES become dominant in terms of  ionizing photons flux. Since their integrated radiation field is much harder than that of OB stars, the heating of the eDIG becomes more efficient. 

We have constructed a finely meshed grid of photoionization models in the framework of this scenario and, for each value of the distance $z$ to the galactic plane, we have selected the ones that fit the observed values of the \oiii/\Hb, \oii/\Hb, and \nii/\Ha\ ratios. Since we have no a priori knowledge of the chemical composition of the eDIG, we had to leave O/H and N/O as free parameters. 

We found that solutions exist for each value of $z$. As a matter of fact, the problem is  not fully determined, since solutions can be obtained for almost any value of O/H.  The models clearly indicate a systematic decrease of the electron density with increasing $z$. If we restrict to solar metallicity, we find that the models which fit the observations become dominated by the HOLMES as $z$ increases. Turning the argument around, this might be an indication that the metallicity of the eDIG is roughly solar. Our models also indicate that N/O increases with increasing distance to the galactic plane, at least until $|z|$$\sim$ 1.5 kpc. This result is in agreement with what common sense would predict about the enrichment of the eDIG by HOLMES progenitors. 

In summary, we have provided a quantitative model to explain the behaviour of the emission lines in the eDIG of NGC 891 using what is known of the populations of massive stars and HOLMES in this galaxy. This galaxy was chosen because it is the one with the largest set of observational data, allowing us to constrain the model relatively well. However, the model being at the same time quite natural and robust, it is likely that it can be applied to the eDIG of other galaxies as well.
 
Obvious next steps of this investigation would be to obtain deep 3D-spectroscopy of extraplanar gas with integral field units, in order to constitute a complete and reliable data set \mycomm{over the entire region of interest.  This will allow us to map the characteristics of the stellar population using a stellar population synthesis code such as  STARLIGHT \citep[see][]{2005MNRAS.358..363C}, and estimate directly the ionizing radiation fields from the HOLMES population, in a way similar to what was done in \citet{2008MNRAS.391L..29S}. This radiation field will then be fed into a  3D photoionization code, such as MOCASSIN \citep{2003MNRAS.340.1153E} which can account for the spatial distribution of the ionizing sources. By confronting the computed emission line ratios with those observed across the field of NGC 891, this should allow us to test in much more detail the scenario proposed here.}
 
\section*{Acknowledgements}

This work is partly supported by grants PAPIIT IN123309 from DGAPA (UNAM,Mexico), CONACyT-49737 and 50296 (Mexico), and CNRS-CONACyT bi-lateral collaboration J010/164/10. 
This paper was first submitted to ApJ. The second referee of that journal made insightful comments on our paper, for which we are grateful. For unclear reasons, we were not allowed to resubmit to ApJ, but that referee's comments are adressed in the present version of the paper. We wish to thank the referee from MNRAS for nice comments and further suggestions.




\appendix

\section{The effect of dust in the eDIG}\label{sec:effect-dust} 
\begin{figure}
\includegraphics[angle=0,width=8cm]{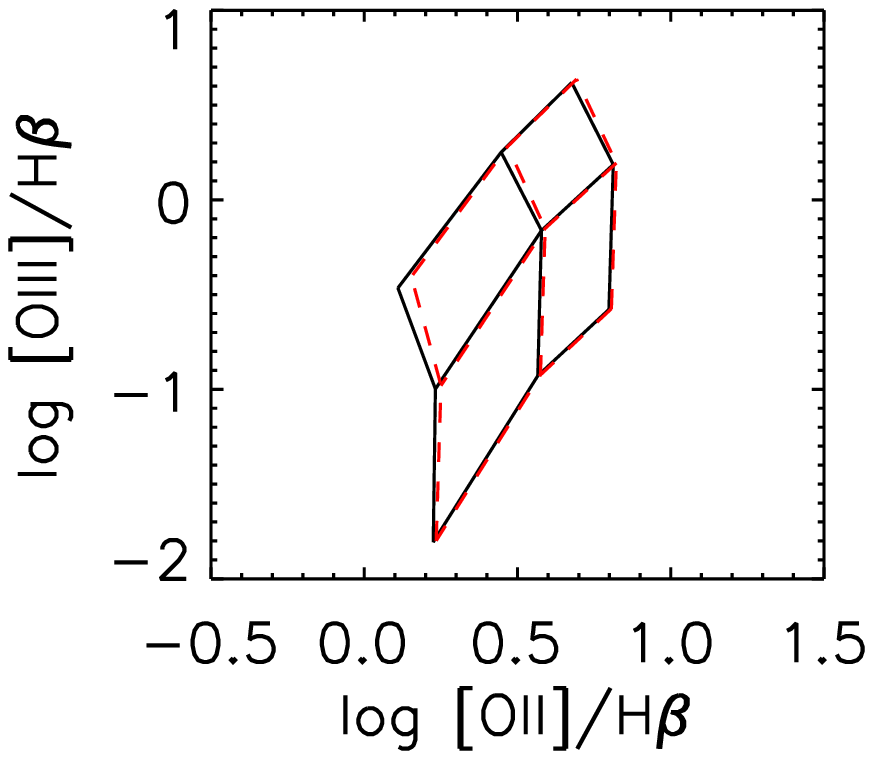}
\caption{The values of  \oiii/\Hb\ vs \oii/\Hb\ for grids of models with and without dust in the eDIG. The models are constructed with log O/H $= -3.3$, log N/H$=-3.8$, log $U =-4., -3.5,$ and $-3.$, and \phiob $ = 4., 5.,$ and 7. Red dotted lines join the dusty models while continuous black lines join the dust-free models.}\label{fig:fig-appen1}
\end{figure}

The eDIG certainly contains dust. In the models presented in this paper, however, dust was not explicitly included, mainly to speed up the computations by concentrating on the basic parameter of the problem, which is the spectral energy distribution of the ionizing radiation field. We however depleted the abundances of metals (Mg, Si and Fe) in all our calculations, as mentioned in Sect. \ref{sec:Cloudy}. As shown by \citet{1995ApJ...454..807S}, it is depletion which introduces the strongest perturbation due to dust in the optical line spectra. The effects linked to the physics of dust (selective absorption, heating and cooling by dust grains) only play a minor role in our problem, as illustrated with the reduced grid of dusty models shown in Fig.  \ref{fig:fig-appen1}. Those models, joined by the red dashed lines, were computed taking the default Cloudy values of dust-to-gas ratio and dust composition, and cover the full range of $U$ and \phiob\ values. For comparison, we show the corresponding dust-free models joining them by continuous black lines. 

\section{The effect of the geometrical distribution of the HOLMES radiation field}
\label{sec:effect-geom-distr}
\begin{figure}
\includegraphics[angle=0,width=8cm]{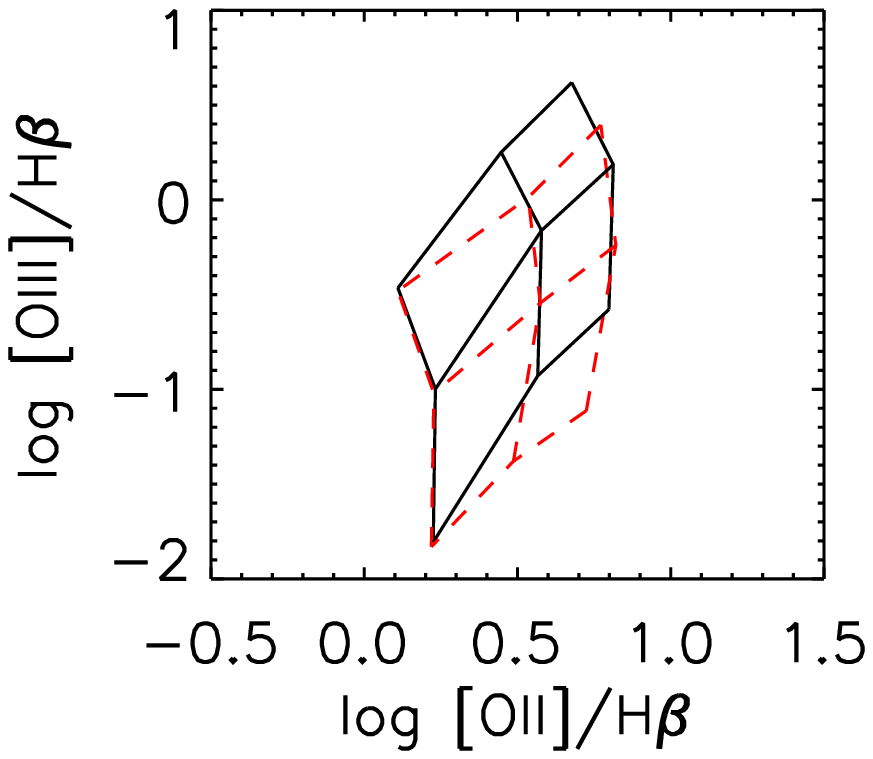}\caption{The values of  \oiii/\Hb\ vs \oii/\Hb\ for grids of models with different treatment of the HOLMES radiation field (see text). The black lines correspond to the same models as in  Fig.~\ref{fig:fig-appen1}, while the red dashed lines correspond to models in which the HOLMES radiation field has been separated into two components, ionizing a DIG cloud from ``above'' and from ``below''(see text).}
\label{fig:fig-appen2}\end{figure}

In the main body of this paper, in order to reduce the parameter space, the transfer of radiation was computed in a simplified way,
assuming that each cloud is ionized from the bottom by the radiation field from both the OB stars and the HOLMES. To show how the effect of
a different geometrical distribution of the HOLMES affects the results, we plot in Fig. \ref{fig:fig-appen2} the values of \oiii/\Hb\ vs \oii/\Hb\ for the same grid of models as shown in the central panel of Fig. \ref{fig:O2O3}. They are joined by continuous (black) lines. We then consider pairs of models constructed with the same values of density and \phiob, but now distributing the radiation of the HOLMES into 0.5\phiholmes from ``below'',  added to the radiation field from the OB stars, and 0.5\phiholmes from ``above''.  Those summed models are joined with (red) dashed lines. Figure \ref{fig:fig-appen2} shows that all the models with the same value of \phiob are found along the same curves, whatever the distribution of the HOLMES radiation. This is because what counts most, in the direction of increasing \oiii/\Hb\ \textit{and } \oii/\Hb, is the global energy provided by the ionizing photons. For a given observational point, however, the model best fitting the data will have a slightly different value of density than
the simple models we considered in the main body of the paper. To obtain similar line ratios with different models, a similar ionization
parameter $U$ is needed, which is achieved by changing the gas density in the same proportions as the photon density. Therefore, the values of $n$ corresponding to the observations are likely to be slightly different from the ones shown in Fig. \ref{fig:resU}, but their variation with $|z|$ should not change, and the main result of our paper, i.e. that the HOLMES expected to be present in the galaxy can explain the observed intensities of the emission lines above the galactic plane, remains unaffected.

\section{The effect of partial absorption of the ionizing radiation from the massive OB stars}\label{sec:effect-tau}

\begin{figure}
\includegraphics[angle=0,width=7cm]{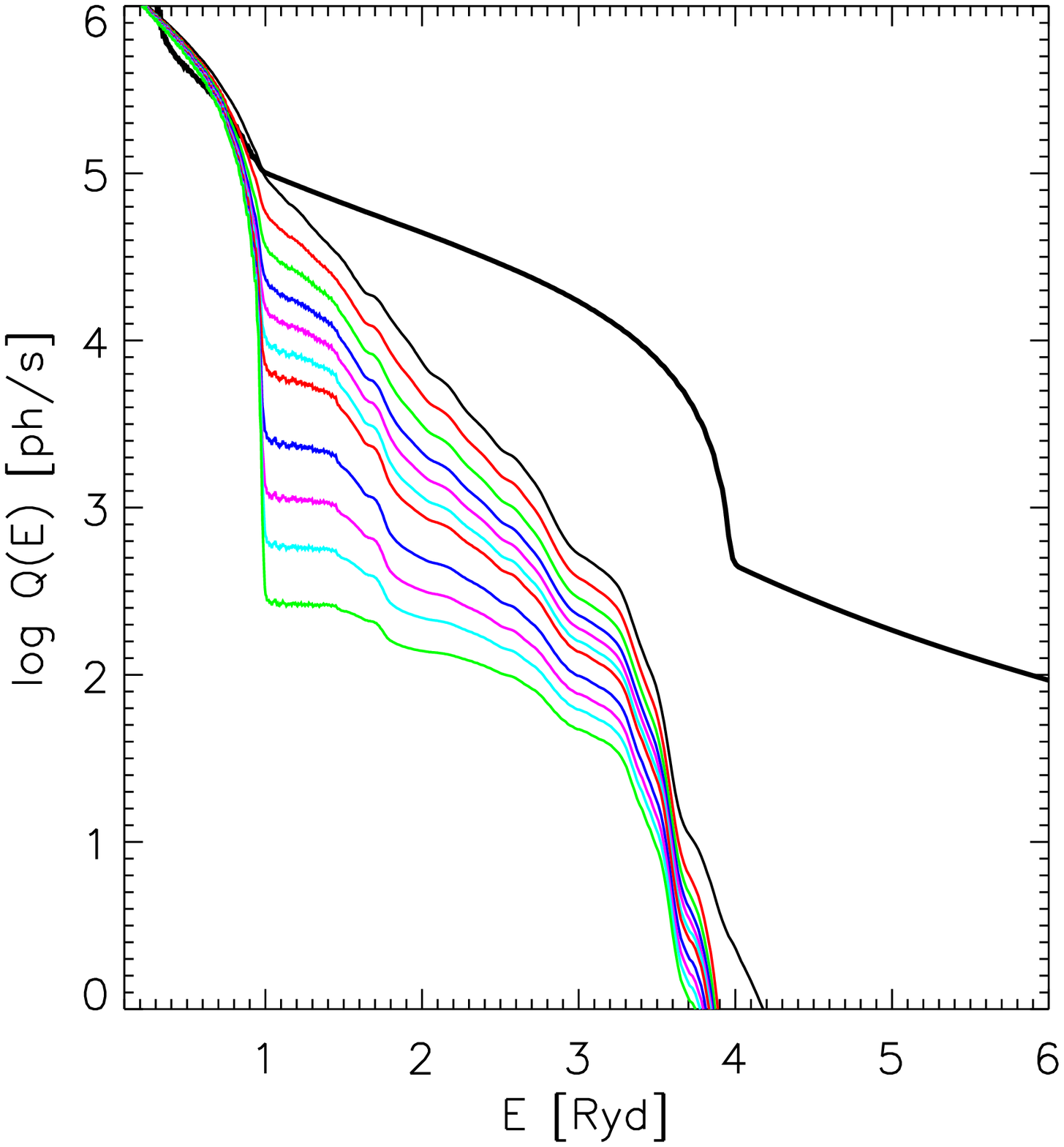}
\caption{The total number, $Q(E)$, of photons above the energy E, as a function of E, normalized to the same value $Q(0)$, leaking out of photoionization models computed with log $U$= -3.5, for different values of the optical thickness at the Lyman edge, $\tau_{13.6}$. The colors (black, red, green, blue, magenta, cyan, red, blue, magenta, cyan and green) correspond to  $\tau_{13.6} = $ 0., 1.15, 2.30, 3.45, 4.6, 5.75, 6.90, 10., 12.9, 15.8, and 20 respectively, for the OB stars. The thick black line shows, for reference, the value of $Q(E)$ for the HOLMES radiation field.} \label{fig:hardening}
\end{figure}

\begin{figure}
\includegraphics[angle=0,width=7cm]{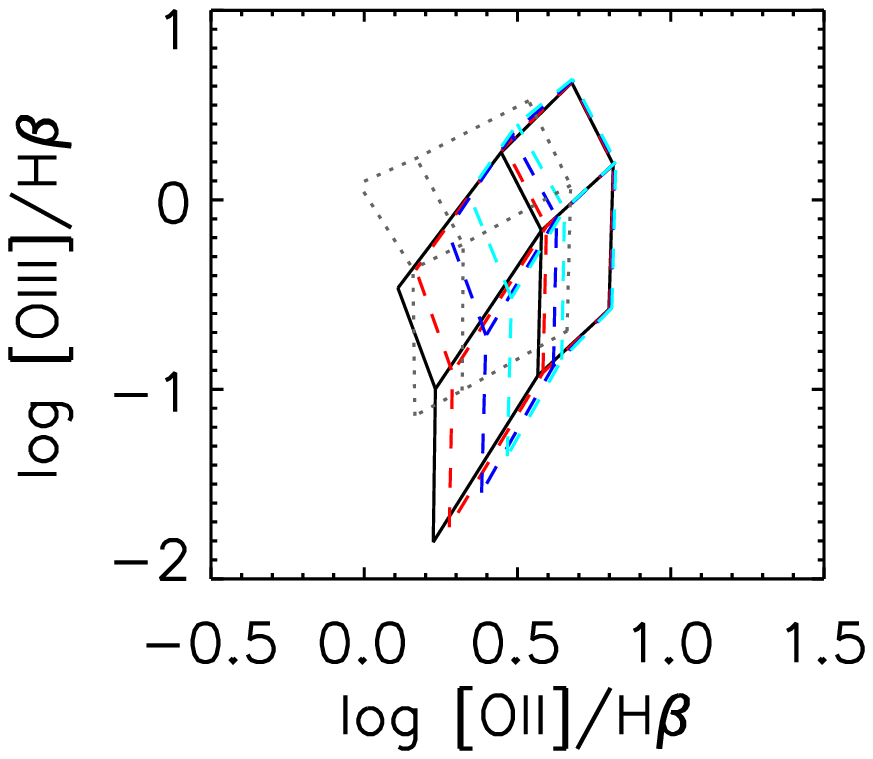}
\caption{The values of  \oiii/\Hb vs \oii/\Hb for grids of models with different values of the partial absorption of the OB radiation field.  The colors (black, red, blue, cyan, and green) correspond to  $\tau_{13.6} = $ 0., 1.15, 3.45, 5.75, and 20 respectively. The models have the same values of O/H, N/O, \phiholmes/\phitot and $U$ as in Fig. \ref{fig:fig-appen1}.}
\label{fig:fig-appen3a}\end{figure}

\paragraph*{The dust-free case}
As mentioned in Sect. \ref{sec:OB}, the radiation field from the OB stars reaching the eDIG clouds may have been affected by  partial absorption, especially in the disk. Figure~\ref{fig:hardening} shows how the values of $Q(E)$, the total number of photons above the energy E, vary with  E, for various values of $\tau_{13.6}$. The absorption produces a hardening of the ionizing radiation. This leads to an increase of the electron temperature in the ionized gas and an enhancement of collisionaly excited lines with respect to recombination lines. 
Figure~\ref{fig:fig-appen3a} shows the effect of this absorption of the OB stellar radiation field on the same reduced grid of photoionization models as shown in black in Fig.~\ref{fig:fig-appen1}, for some of the values of $\tau_{13.6}$ considered in Fig. ~\ref{fig:hardening} (and adopting the same colours). We see that the effect of this hardening on our photoionization models, while appreciable at low values of \phiholmes/\phitot,  becomes  negligible when \phiholmes/\phitot exceeds 50\%. This justifies our neglecting the effect of partial absorption in our high resolution grid used in Sect. \ref{sec:results}, since we are mainly interested in explaining the line ratios of the eDIG at high values of $|z|$. 

\begin{figure*}
\includegraphics[angle=0,width=14cm]{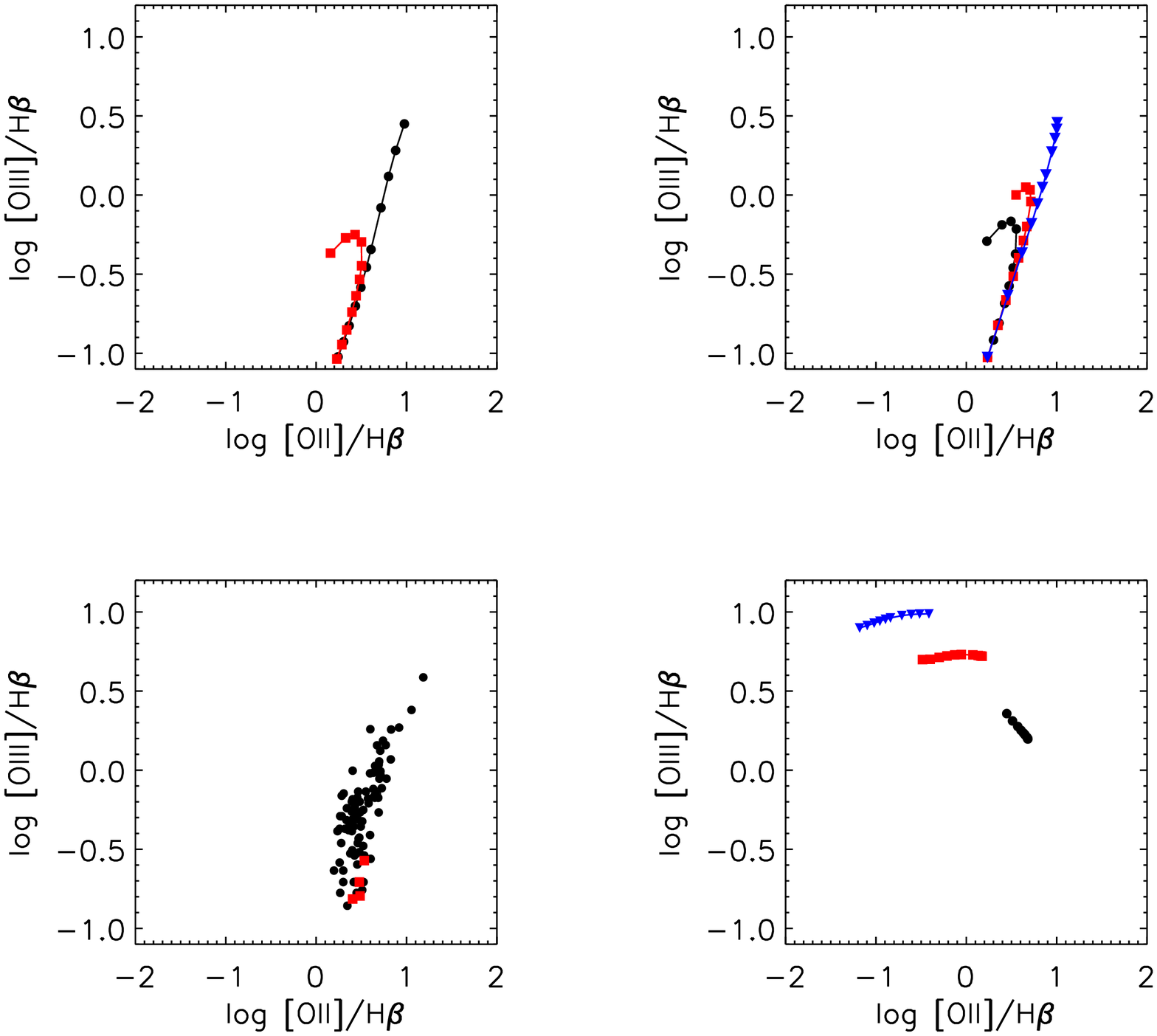}
\caption{\oiii/\Hb\ vs \oii/\Hb. \textbf{Upper left:} Models of the eDIG ionized by radiation from OB stars (no HOLMES) that suffered partial absorption by \textit{dust-free }\hii\ regions in the disk. The models have the same values of O/H,  and N/O  as in Fig. \ref{fig:fig-appen1},  and log $U$=-3.5. The symbols mark different values of the partial absorption ($\tau_{13.6} = $ 0., 1.15, 2.30, 3.45, 4.6, 5.75, 6.90, 10., 12.9, 15.8, and 20 from bottom to top). The black line with filled circles connects models where fluorescence is disabled, while the red line with square symbols connects models with fluorescence taken into account. \textbf{Upper right:} Models of the eDIG ionized by radiation from OB stars (no HOLMES) that suffered partial absorption by \textit{dusty} \hii\ regions in the disk. Fluorescence is taken into account. The black, red and blue curves correspond to absorbers with  ionization parameters log $U$ equal to -3, -2, and -1 respectively. The symbols (circles, squares and triangles, respectively) mark the same values of the partial absorption as in the upper left panel, but this time $\tau_{13.6}$ includes absorption by dust grains. \textbf{Lower left: }The observations from \citet{2001ApJ...560..207O}. Those corresponding to $|z| < 0.3$\,kpc are marked in red, the others in black. \textbf{Lower right:}  Calculated emission in the disk from \hii\ regions responsible for the absortion of the ionizing continuum used by the eDIG models shown in the upper right panel. The colors and symbols are the same (except the meaningless case $\tau_{13.6}$=0, which is not shown).}
\label{fig:fig-appen3b}
\end{figure*}

The grey dotted lines in Fig. \ref{fig:fig-appen3a} correspond to $\tau_{13.6} =20$. For this value of  $\tau_{13.6}$, the hardness of the partially absorbed OB radiation field is equivalent to that of the HOLMES, as can be seen in Fig. \ref{fig:hardening}, so that one might think that the predicted line ratios would be very similar to those of models photoionized exclusively by HOLMES.  Interestingly, Fig. \ref{fig:fig-appen3a} shows that this is not the case. This is due to the effect of fluorescent excitation of \Hb\ by stellar photons around the Lyman lines, which are numerous with respect to ionizing photons in such a circumstance, as can be seen in Fig.~\ref{fig:hardening}, leading to a decrease of the \oii/\Hb\ and \oiii/\Hb ratios due to a strong increase of \Hb. 
The effect of fluorescence  is better illustrated by the top left panel of Fig.~\ref{fig:fig-appen3b}, where the dependence of the  \oiii/\Hb\ and \oii/\Hb\ ratios with respect to  $\tau_{13.6}$ is shown. A single set of parameters (O/H, N/O, \phiob, and $U$) is used. Two cases are shown: the red curve with square symbols corresponds to models taking fluorescence into account; the black curve with filled circles corresponds to models without fluorescence. When fluorescence is not taken into account, one can reach the region of highest \oii/\Hb\ and \oiii/\Hb values observed in the eDIG using the hardened radiation from massive OB stars only  (for convenience, the observational points are plotted again in the lower left panel of Fig.~\ref{fig:fig-appen3b}).  This  was the hypothesis of \citet{1993ASPC...35..540S}, but we show here that the high values of the two line ratios are not accessible, due to fluorescence which increases the \Hb\ line intensities. \mycomm{Notice that these models are performed using version c08.01 of Cloudy, which  treats the fluorescence correctly.}

\paragraph*{The dusty case}

Accounting for the presence of dust in the \hii regions of the disk reduces the effect of fluorescence, since dust can absorb the stellar radiation below the 1~Rydberg limit.  In  the upper right panel of Fig.~\ref{fig:fig-appen3b},  we explore the effect of the presence of dust in the \hii\ regions of the disk, taking the default Cloudy values of dust-to-gas ratio and dust composition. The symbols correspond to the same values of $\tau_{13.6}$ as in the upper left panel, but this time we take into account the absorption of both gas and dust. In all the models, fluorescence is taken into account. Each sequence corresponds to different values of the ionization parameter. The black curve marked with circles corresponds to log $U=-3$, the red one with squares to log $U=-2$ and the blue one with triangles to log $U=-1$. These curves mimic different stages of the development of an \hii\ region, with the largest values of $U$ corresponding to the youngest stages, where the gas is more compact and the effects of dust are important. As the \hii\ region ages, it expands, the ionization parameter drops and the effects of dust become less important, even when keeping the same dust-to-gas mass ratio. In the case of a dust dominated attenuation (ie. log $U = -1$), the eDIG models ionized by partially absorbed radiation from massive OB stars can reach the highest values of the observed  \oii/\Hb\ and \oiii/\Hb ratios. However in this case, the  emission line ratios \textit{of the absorber}, which are shown in the bottom right panel of Fig.~\ref{fig:fig-appen3b}, correspond to a zone in the \oiii/\Hb\ vs \oii/\Hb diagram which is very far from what is observed in the disk (represented by the red points in the lower left panel).  This means that \hii\ regions with such high values of $U$ have a negligible contribution to the overall emission in the disk and, therefore, that leakage of ionizing radiation from such  \hii\ regions into the eDIG must be negligible.

\label{lastpage}

\end{document}